\documentclass[
    ,final            
  ]
  {aipproc}

\layoutstyle{8x11single}

\begin{document}

\title{The Emergent Universe scheme and Tunneling}


\author{Pedro Labra\~{n}a}{
  address={Departamento de F\'{\i}sica, Universidad del
B\'{\i}o-B\'{\i}o, Avenida Collao 1202, Casilla 5-C, Concepci\'on,
Chile and
Departament d'Estructura i Constituents de la Mat\`{e}ria and
Institut de Ci\`{e}ncies del Cosmos, Universitat de Barcelona,
Diagonal 647, 08028 Barcelona, Spain.} }

\begin{abstract}
 We present an alternative scheme for an Emergent Universe scenario,
developed previously in Phys.\ Rev.\ D {\bf 86}, 083524 (2012),
where the universe is initially in a static state supported by a
scalar field located in a false vacuum. The universe begins to
evolve when, by quantum tunneling, the scalar field decays into a
state of true vacuum.
The Emergent Universe models are interesting since they provide
specific examples of non-singular inflationary universes.
\end{abstract}

\maketitle


\section{Introduction}
\label{Int}

Cosmological inflation has become an integral part of the standard
model of the universe. Apart from being capable of removing the
shortcomings of the standard cosmology, it gives important clues for
large scale structure formation \cite{Guth1, Albrecht, Linde1,
Linde2} (see \cite{libro} for a review).
The scheme of inflation is based on the idea that there was an early
phase, before the Big Bang, in which the universe evolved through a
nearly exponential expansion during a short period of time at high
energy scales. During this phase, the universe was dominated by a
potential of a scalar field, which is called the inflaton.
%


In this context, singularity theorems have been devised that apply
in the inflationary scenario, showing that the universe necessarily
had a beginning \cite{Borde:1993xh,Borde:1997pp, Borde:2001nh,
Guth:1999rh,Vilenkin:2002ev}.
However, recently, models that escape this conclusion has been
studied in Refs.
\cite{Ellis:2002we,Ellis:2003qz,Mulryne:2005ef,Mukherjee:2005zt,
Mukherjee:2006ds,Banerjee:2007qi,Nunes:2005ra,Lidsey:2006md}. These
models, called Emergent Universe (EU), do not satisfy the
geometrical assumptions of these theorems. Specifically, the
theorems assume that either {\bf i)} the universe has open space
sections, implying $k = 0$ or $-1$, or \textbf{ii)} the Hubble
expansion rate $H$ is bounded away from zero in the past, $H > 0$.

Normally in the Emergent Universe scenario, the universe is
positively curved and initially it is in a past eternal classical
Einstein static state which eventually evolves into a subsequent
inflationary phase, see
\cite{Ellis:2002we,Ellis:2003qz,Mulryne:2005ef,Mukherjee:2005zt,
Mukherjee:2006ds,Banerjee:2007qi,Nunes:2005ra,Lidsey:2006md}.

\begin{figure}
\centering
\includegraphics[width=10cm]{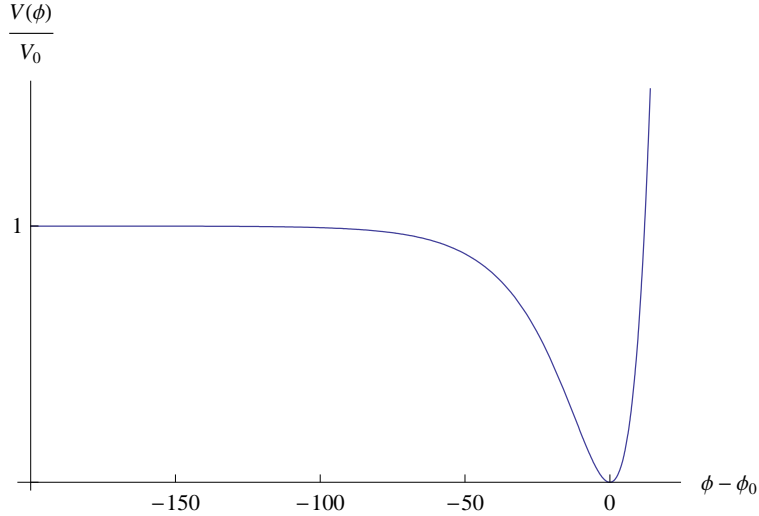}
\caption{Schematic representation of a potential for a standard
Emergent Universe scenario. \label{Potencial-1}}
\end{figure}

For example, in the original scheme \cite{Ellis:2002we,
Ellis:2003qz}, it is assumed that the universe is dominated by a
scalar field (inflaton) $\phi$ with a scalar potential $V(\phi)$
that approach a constant $V_0$ as $\phi \rightarrow -\infty$ and
 monotonically rise once  the scalar field exceeds a
certain value $\phi_0$, see Fig. (\ref{Potencial-1}).

During the past-eternal static regime it is assumed that the scalar
field is rolling on the asymptotically flat part of the scalar
potential with a constant velocity, providing the conditions for a
static universe. But once the scalar field exceeds some value, the
scalar potential slowly droops from its original value. The overall
effect of this is to distort the equilibrium behavior breaking the
static solution.
If the potential has a suitable form in this region, slow-roll
inflation will occur, thereby providing a 'graceful entrance' to
early universe inflation.


Notice that, as was shown by Eddington \cite{Eddington}, the
Einstein static state is unstable to homogeneous perturbations. This
situation has implication for the Emergent Universe scenario, see
discussion in Sec. \ref{Sec-conclusions}.


This scheme for a Emergent Universe have been used not only on
models based on General Relativity \cite{Ellis:2002we,
Ellis:2003qz}, but also on models where non-perturbative quantum
corrections of the Einstein field equations are considered
\cite{Mulryne:2005ef,Nunes:2005ra, Lidsey:2006md}, in the context of
a scalar tensor theory of gravity \cite{delCampo:2007mp,
delCampo:2009kp} and recently in the framework of the so-called two
measures field theories \cite{delCampo:2010kf,delCampo:2011mq,
Guendelman:2011fq,Guendelman:2011fr, review-guendelman}.
%


Another possibility for the Emergent Universe scenario is to
consider models in which the scale factor asymptotically tends to a
constant in the past \cite{Mukherjee:2005zt, Mukherjee:2006ds,
Banerjee:2007sg,Debnath:2008nu, Paul:2008id, Beesham:2009zw,
Debnath:2011qi, Mukerji:2011wq}.


The Emergent Universe models are appealing since they provide
specific examples of non–singular (geodesically complete)
inflationary universes. Furthermore, it has been proposed that
entropy considerations favor the ES state as the initial state for
our universe \cite{Gibbons:1987jt,Gibbons:1988bm}.

Also, it has been proposed  \cite{Labrana:2013oca} that the super
inflation phase, which is a characteristic shared by all EU models,
could be responsible for part of the anomaly in the low multipoles
of the CMB, in particular to the observed lack of power at large
angular scales \cite{cobe, WMAP1,planckI, PlanckXXII, PlanckXV,
PlanckXXIII}.


We can note that both schemes for a Emergent Universe are not truly
static during the static regime. For instance, in the first scheme
during the static regime the scalar field  is rolling on the flat
part of its potential. On the other hand, for the second scheme the
scale factor is only asymptotically static.

\begin{figure}
\centering
\includegraphics[width=10cm]{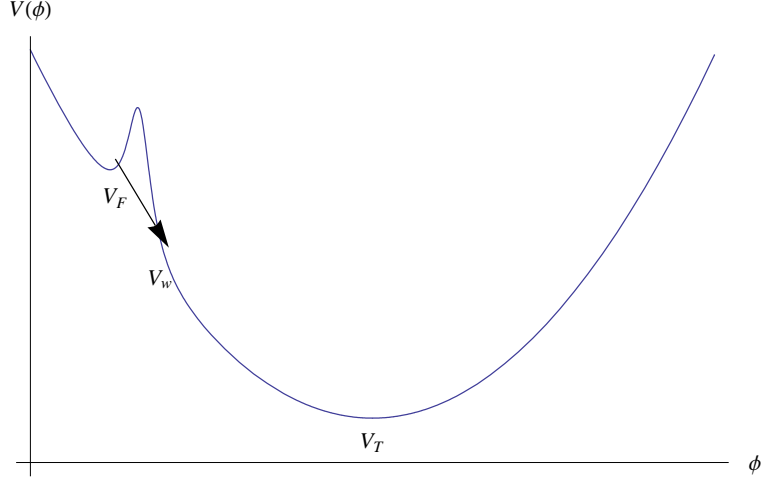}
\caption{A double-well inflationary potential $V(\phi)$. In the
graph, some relevant values  are indicated. They are the false
vacuum $V_F=V(\phi_F)$ from which the tunneling begins,
$V_W=V(\phi_W)$ where the tunneling stops and where the inflationary
era begins, while $V_T=V(\phi_T)$ denote the true vacuum energy.
\label{Potential-2}}
\end{figure}

However, recently, it has been proposed an alternative scheme for an
Emergent Universe scenario, where the universe is initially  in a
truly static state \cite{Labrana:2011np}. This state is supported by
a scalar field which is located in a false vacuum ($\phi = \phi_F$),
see Fig.(\ref{Potential-2}). The universe begins to evolve when, by
quantum tunneling, the scalar field decays into a state of true
vacuum. Then, a small bubble of a new phase of field value $\phi_W$
can form, and expand as it converts volume from high to low vacuum
energy and feeds the liberated energy into the kinetic energy of the
bubble wall \cite{Coleman:1977py, Coleman:1980aw}. Inside the
bubble, space-like surfaces of constant $\phi$ are homogeneous
surfaces of constant negative curvature. One way of describing this
situation is to say that the interior of the bubble always contains
an open Friedmann-Robertson-Walker universe \cite{Coleman:1980aw}.
If the potential has a suitable form, inflation and reheating may
occur in the interior of the bubble as the field rolls from $\phi_W$
to the true minimum at $\phi_T$, in a similar way to what happens in
models of Open Inflationary Universes, see for example \cite{linde,
re8, delC1, delC2, Balart:2007je}.


\begin{figure}
\centering
\includegraphics[width=10cm]{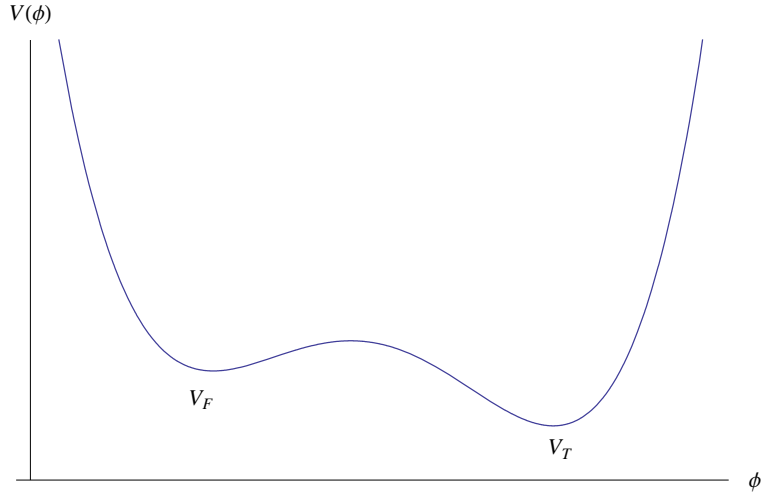}
\caption{Potential with a false and true vacuum.
\label{Potential-3}}
\end{figure}

In Ref.~\cite{Labrana:2011np} we considered a simplified version of
this scheme, where we focused on studied the process of creation and
evolution of a bubble of true vacuum in the background of an ES
universe.
In particular, we considered an inflaton potential similar to Fig.
(\ref{Potential-3}) and studied the process of tunneling of the
scalar field from the false vacuum $\phi_F$ to the true vacuum
$\phi_T$ and the consequent creation and evolution of a bubble of
true vacuum in the background of an ES universe.
Here we review the principal results of Ref.~\cite{Labrana:2011np}.

\section{Static Universe Background}\label{Sec-static}

Based on the standard Emergent Universe (EU) scenario, we consider
that the universe is positively curved and it is initially in a past
eternal classical Einstein static state. The matter of the universe
is modeled by a standard perfect fluid $P=(\gamma -1)\rho$ and a
scalar field (inflaton) with energy density $\rho_\phi =
\frac{1}{2}(\partial_t\phi)^2 + V(\phi)$ and pressure $P_\phi =
\frac{1}{2}(\partial_t\phi)^2 - V(\phi)$. The scalar field potential
$V(\phi)$ is depicted in Fig.~\ref{Potential-3}. The global minimum
of $V(\phi)$ is tiny and positive, at a field value $\phi_T$, but
there is also a local false minimum  at $\phi=\phi_F$.

The metric for the static state is given by the closed
Friedmann-Robertson-Walker metric:

\begin{equation}
d{s}^{2}\,=\, d{t}^{2}\,-\,
a(t)^{2}\left[\frac{dr^2}{1-\frac{r^2}{R^2}} + r^2\,(d\theta^2 +
\sin^2 \!\!\theta\,\, d\phi^2 )\right], \label{met1} \end{equation}
where $a(t)$ is the scale factor, $t$ represents the cosmic time and
the constant $R>0$.
We have explicitly written $R$ in the metric in order to make more
clear the effects of the curvature on the bubble process
(probability of creation and propagation of the bubble).

Given that there are no interactions between the standard fluid and
the scalar field, they separately obey energy conservation and
Klein– Gordon equations,
\begin{eqnarray}
\partial_t \rho + 3\gamma \,H\,\rho = 0 \label{rho}\,,\\
\nonumber \\
\partial^2_t\phi+3H\,\partial_t\phi=-\frac{\partial
V(\phi)}{\partial\phi}\,, \label{phi}
\end{eqnarray}
where $H = \partial_t a /a$.

The Friedmann and the Raychaudhuri field equations become,
\begin{eqnarray}
H^2 = \frac{8\pi G}{3}\left(\rho + \frac{1}{2}(\partial_t\phi)^2 +
V(\phi)
\right) - \frac{1}{R^2 a^2}, \label{H} \\
\nonumber\\
\partial^2_t a = - \frac{8\pi
G}{3}\,a\left[\left(\frac{3}{2}\gamma -1\right)\rho + \dot{\phi}^2 -
V(\phi) \right]\label{H2}.
\end{eqnarray}

The  static universe is characterized by the conditions
$a=a_0=const.$, $\partial_ta_0=\partial^2_ta_0=0$ and
$\phi=\phi_F=Cte.$, $V(\phi_F) = V_F$ corresponding to the false
vacuum.

From Eqs.~(\ref{rho}) to (\ref{H2}), the static solution for a
universe dominated by a scalar field placed in a false vacuum and a
standard perfect fluid, are obtained if the following conditions are
met
\begin{eqnarray}
\rho_0 &=& \frac{1}{4\pi G}\,\frac{1}{\gamma\,R^2 a^2_0}\,, \\
\nonumber \\
 V_F &=& \left(\frac{3}{2}\gamma -1\right)\rho_0 \,,
\end{eqnarray}
where $\rho_0$ is energy density of the perfect fluid present in the
static universe. Note that $\gamma > 2/3$ in order to have a
positive scalar potential.

By integrating Eq.~(\ref{rho}) we obtain
\begin{equation}
\rho = \frac{A}{a^{3\gamma}}\,,
\end{equation}
where $A$ is an integration constant.
By using this result, we can rewrite the conditions for a static
universe as follow
\begin{eqnarray}
A = \frac{1}{4\pi G}\;\frac{a^{3\gamma-2}_0}{\gamma\,R^2}\,,\\
\nonumber \\
V_F = \left(\frac{3}{2}\gamma -1\right)\frac{1}{4\pi
G}\,\frac{1}{\gamma\,R^2 a^2_0}\,.
\end{eqnarray}

In a purely classical field theory if the universe is static and
supported by the scalar field located at the false vacuum $V_F$,
then the universe remains static forever. Quantum mechanics makes
things more interesting because the field can tunnel through the
barrier and by this process create a small bubble where the field
value is $\phi_T$. Depending of the background where the bubble
materializes, the bubble could expanded or collapsed
\cite{Fischler:2007sz, Simon:2009nb}.

\section{Bubble Nucleation}\label{Sec-Bubble}

In this section we study the tunneling process of the scalar field
from the false vacuum to the true vacuum and the consequent creation
of a bubble of true vacuum in the background of Einstein static
universe.
Given that in our case the geometry of the background correspond to
a Einstein static universe and  not a de Sitter space, we proceed
following the scheme developed in
\cite{KeskiVakkuri:1996gn,Simon:2009nb}, instead of the usual
semiclassical calculation of the nucleation rate based on instanton
methods \cite{Coleman:1980aw}.
In particular, we will consider the nucleation of a spherical bubble
of true vacuum $V_T$   within the false vacuum $V_F$. We will assume
that the layer which separates the two phases (the wall) is of
negligible thickness compared to the size of the bubble (the usual
thin-wall approximation). The energy budget of the bubble consists
of latent heat (the difference between the energy densities of the
two phases) and surface tension.

In order to eliminate the problem of  predicting the reaction of the
geometry to an essentially a-causal quantum jump, we neglect during
this computation the gravitational back-reaction of the bubble onto
the space-time geometry.

The gravitational back-reaction of the bubble will be consider in
the next chapter when we study the evolution of the bubble after its
materialization.

In our case the shell trajectory  follows from the action (see
\cite{KeskiVakkuri:1996gn, Basu:1991ig})
\begin{equation}\label{accion}
S = \int dy \Bigg\{ 2\pi\,\epsilon\, \bar{a}_0^4\Big[\chi -
\cos(\chi)\sin(\chi)\Big] - 4\pi\,\sigma \,\bar{a}_0^3
\,\sin^2(\chi)\sqrt{1- {\chi'}^2}\Bigg\}.
\end{equation}
where we have denoted the coordinate radius of the shell as $\chi$,
and we have written the static ($a = a_0 = Cte.$) version of the
metric Eq.(\ref{met1}) as
\begin{equation}
\label{met2} ds^2 = \bar{a}_0^2\Big(dy^2 - d\chi^2 -
\sin^2(\chi)d\Omega^2 \Big),
\end{equation}
with $\frac{r}{R} = \sin(\chi)$, $\bar{a}_0 = R\,a_0$, $dt =
\bar{a}_0\,dy$ and prime means derivatives respect to $y$.

In the action  (\ref{accion}), $\epsilon$ and $\sigma$ denote,
respectively, the latent heat and the surface energy density
(surface tension) of the shell.

The action (\ref{accion}) describes the classical trajectory of the
shell after the tunneling. This trajectory emanates from a classical
turning point, where the canonical momentum
\begin{equation}
\label{PP} P = \frac{\partial S}{\partial \chi'} = 4\pi\,\sigma\,
\bar{a}_0^3\,\chi'\,\frac{\sin^2(\chi)}{\sqrt{1- {\chi'}^2}}\,,
\end{equation}
vanishes \cite{KeskiVakkuri:1996gn}.
In order to consider tunneling, we evolve this solution back to the
turning point, and then try to shrink the bubble to zero size along
a complex $y$ contour, see \cite{KeskiVakkuri:1996gn,Simon:2009nb}.
For each solution, the semiclassical tunneling rate is determined by
the imaginary part of its action, see \cite{KeskiVakkuri:1996gn}:
\begin{equation}
\label{P} \Gamma \approx e^{-2Im [S]}\,.
\end{equation}

From the action (\ref{accion}) we found the equation of motion
\begin{equation}
\label{ecmov} \frac{\sin^2(\chi)}{\sqrt{1-{\chi'}^2}} =
\frac{\epsilon \, \bar{a}_0}{2 \sigma} \Big[\chi -
\cos(\chi)\sin(\chi)\Big].
\end{equation}

The action (\ref{accion}) can be put in a useful form by using
Eq.(\ref{ecmov}), and changing variables to $\chi$:

\begin{equation}
S = \int d\chi
\,\frac{4\pi}{3}\,\epsilon\,a_0^4\,\sin^2(\chi)\sqrt{\left(\frac{3[\chi
- \cos(\chi)\sin(\chi)]}{2\sin^2(\chi)}\right)^2 -\, \bar{r}_0^2}
\;,
\end{equation}
where $\bar{r}_0 = \frac{r_0}{R}$ and $r_0 = \frac{3
\sigma}{\epsilon \,a_0}$ is the radio of nucleation of the bubble
when the space is flat ($R \rightarrow \infty)$ and static (i.e.
when the space is Minkowsky).

The nucleation radius $\bar{\chi}$ (i.e. the coordinate radius of
the bubble at the classical turning point), is a solution to the
condition $P =0$. Then from Eq.~(\ref{PP}) we obtain
\begin{equation}
\label{radio} \frac{\bar{\chi} -
\cos(\bar{\chi})\sin(\bar{\chi})}{\sin^2(\bar{\chi})} = \frac{2
\sigma}{\epsilon \, \bar{a}_0}.
\end{equation}

The action (\ref{accion}) has an imaginary part coming from the part
of the trajectory $0 < \chi < \bar{\chi}$, when the bubble is
tunneling:
\begin{equation}\label{im-s}
Im[S] = \frac{4\pi}{3}\,\epsilon\,a_0^4\,\int^{\bar{\chi}}_0 d\chi
\,\sin^2(\chi)\sqrt{\bar{r}_0^2 -\, \left(\frac{3[\chi -
\cos(\chi)\sin(\chi)]}{2\sin^2(\chi)}\right)^2} \;,
\end{equation}

Expanding (\ref{im-s}) at first nonzero contribution in $\beta=
(r_0/R)^2$ we find
\begin{equation}
\label{im-s1} Im[S] = \frac{27\,\sigma^4\,\pi}{4\,\epsilon^3}\Big[1
- \frac{1}{2}\beta^2 \Big]
\end{equation}

This result is in agreement with the expansion obtained in
\cite{Abbott:1987xq}.
Then, the nucleation rate is
\begin{equation}
\Gamma \approx e^{-2Im S} \approx
 \exp\left[-\frac{27 \sigma^4 \pi}{2
\epsilon^3} \left(1 - \frac{9\sigma^2}{2\epsilon^3\,a_0^2R^2}
\right) \right].
\end{equation}

We can note that the probability of the bubble nucleation is
enhanced by the effect of the curvature of the closed static
universe background.

\section{Evolution of the Bubble}\label{Sec-evolution}

In this section we study the evolution of the bubble after the
process of tunneling.
During this study we are going to consider the gravitational
back-reaction of the bubble.
We follow the approach used in \cite{Fischler:2007sz} where it is
assumed that the bubble wall separates space-time into two parts,
described by different metrics and containing different kinds of
matter.
The bubble wall is a timelike, spherically symmetric hypersurface
$\Sigma$, the interior of the bubble is described by a de Sitter
space-time  and the exterior by the static universe discussed in
Sec.~\ref{Sec-static}. The Israel junction conditions
\cite{Israel:1966rt} are implement in order to joint these two
manifolds along there common boundary $\Sigma$. The evolution of the
bubble wall is determined by implement these conditions.
Unit as such that $8\pi \,G = 1$.
The exterior of the bubble is described by the metric
Eq.~(\ref{met1}) and the equations (\ref{rho}-\ref{H2}), previously
discussed in Sec.~\ref{Sec-static}. At the end, the static solution
for these equations will be assumed. The interior of the bubble will
be described by the metric of the \mbox{de Sitter} space-time in its
open foliation, see \cite{Coleman:1980aw}

\begin{equation}\label{metrica-interior}
ds^2 = dT^2 - b^2(T)\left( \frac{dz^2}{1+z^2} + z^2\,d\Omega_2
\right),
\end{equation}

where the scale factor satisfies

\begin{equation}
\left(\frac{db}{dT} \right)^2 = \left(\frac{V_T}{3}\right) b^2(T) +1
\,.
\end{equation}

These two regions are separated by the bubble wall $\Sigma$, which
will be assumed to be a thin-shell and spherically symmetric.
Then, the intrinsic metric on the shell is \cite{Berezin:1987bc}
\begin{equation}\label{metrica-burbuja}
ds^2|_\Sigma = d\tau^2 - B^2(\tau)\,d\Omega_2 \,,
\end{equation}
where $\tau$ is the shell proper time.

Now we proceed to impose the Israel matching conditions
\cite{Israel:1966rt} in order to joint the manifolds along there
common boundary $\Sigma$.
The first of Israel's conditions impose that the metric induced on
the shell from the bulk 4-metrics on either side should match, and
be equal to the 3-metric on the shell.
Then by looking from the outside to the bubble-shell we can
parameterize the coordinates  $r = x(\tau)$ and $t = t(\tau)$,
obtaining the following match conditions, see \cite{Fischler:2007sz}

\begin{equation}
a(t)x = B(\tau)\,,\;\;\;\;\;
\left(\frac{dt}{d\tau} \right)^2 = 1 + \frac{a(t)^2}{1-
\left(\frac{x}{R}\right)^2}\left(\frac{dx}{d\tau}\right)^2\,,\label{cond1}
\end{equation}
where all the variables in these equations are thought as functions
of $\tau$.
On the other hand, the  angular coordinates of metrics (\ref{met1})
and (\ref{metrica-burbuja}) can be just identified in virtue of the
spherical symmetry.


The second junction condition could be written as follow
\begin{equation}\label{segunda-condicion}
[K_{ab}] - h_{a b}[K] = S_{ab},
\end{equation}
where $K_{ab}$ is the extrinsic curvature of the surface $\Sigma$
and square brackets stand for discontinuities across the shell.
Following \cite{Fischler:2007sz}, we assume that the surface
energy-momentum tensor $S_{ab}$ has a perfect fluid form given by
${S_\tau}^\tau \equiv \sigma$ and ${S_\theta}^\theta = {S_\phi}^\phi
\equiv -\bar{P}$, where $\bar{P} = (\bar{\gamma} -1)\sigma$.

In the outside coordinates we parameterize $x(t)$ as the curve for
the bubble evolution (the bubble radius in these coordinates). Since
$x$ and $t$ are dependent variables on the shell, this is
legitimate.

Then, from the Israel conditions we can obtain the following
equation for the evolution of $x(t)$ and $\sigma(t)$, see
\cite{Labrana:2011np}

\begin{equation}\label{ecuacion-x}
\frac{dx}{dt} = \pm \sqrt{\frac{\left(R^2-
x^2\right)\left(a_0^2\,C^2\,x^2
-1\right)}{x^2\,a_0^2\left(a_0^2\,C^2\,R^2 - 1\right)}}\,,
\end{equation}

\begin{equation}\label{ecuacion-sigma}
\frac{d\sigma}{dt} =
-2\left(\frac{\bar{\gamma}\,\sigma}{x}\right)\frac{dx}{dt} +
\frac{a_0\,\gamma\,\rho_0}{\sqrt{-\left(\frac{dx}{dt}\right)^2a_0^2
+ 1 -\frac{x^2}{R^2}}}\,\frac{dx}{dt}\,.
\end{equation}

Where
\begin{equation}\label{radio-burbuja2}
C^2 = \frac{V_T}{3} + \left(\frac{\sigma}{4} +
\frac{1}{\sigma}\left[\frac{V_F - V_T}{3} +
\frac{A}{3a^{3\gamma}}\right]\right)^2.
\end{equation}

The positive energy condition $\sigma > 0$ together with Israel
conditions impose the following restriction to $\sigma$
\begin{equation}
 0 < \sigma \leq 2 \sqrt{\frac{V_F-V_T}{3} +
\frac{\rho_0}{3}}\;.
\end{equation}

Also, from the definition of $x$ and Eq.(\ref{ecuacion-x}) we obtain
the following restriction for $x$
\begin{equation}
\frac{1}{a_0C} \leq x \leq R \;.
\end{equation}

We solved the Eqs.~(\ref{ecuacion-x}, \ref{ecuacion-sigma})
numerically by consider different kind and combinations of the
matter content of the background and the bubble wall.
From these solutions we found that once the bubble has materialized
in the background of an ES universe, it grows filling completely the
background space.

In order to find the numerical solutions we  chose the following
values for the free parameters of the model, in units where $8\pi G
= 1$:
\begin{eqnarray}
a_0 &=& 1 \,,\\
V_T &=& 0.1 V_F \,,\\
\sigma_{init} &=& 10^{-6}\,.
\end{eqnarray}

The other parameters are fixed by the conditions discussed in
section two.


\begin{figure}
\centering
\includegraphics[width=7cm]{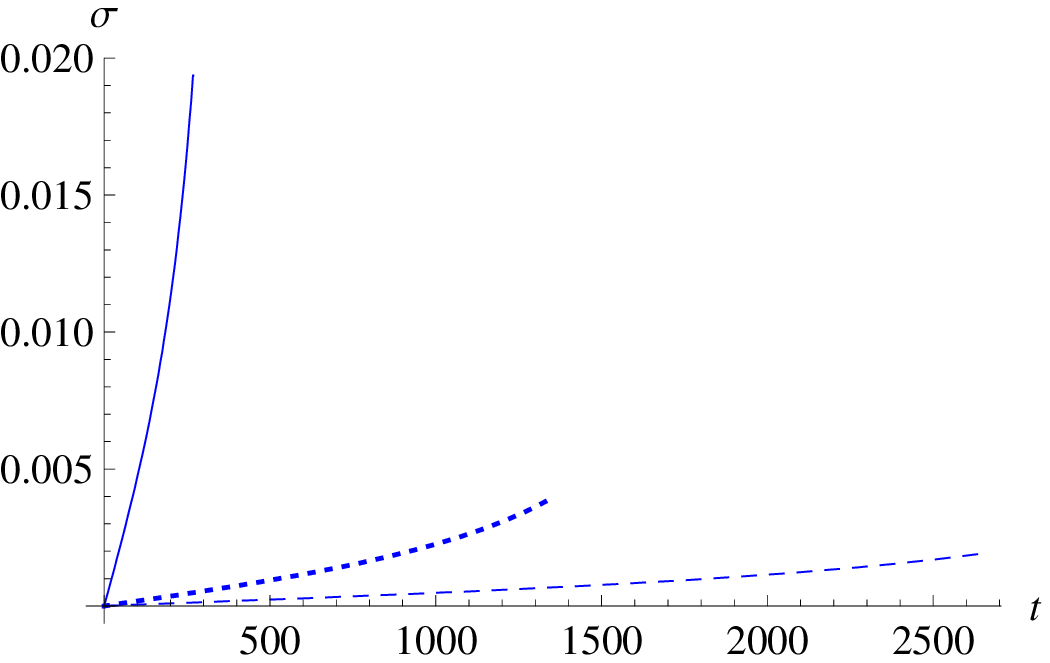}
\includegraphics[width=7cm]{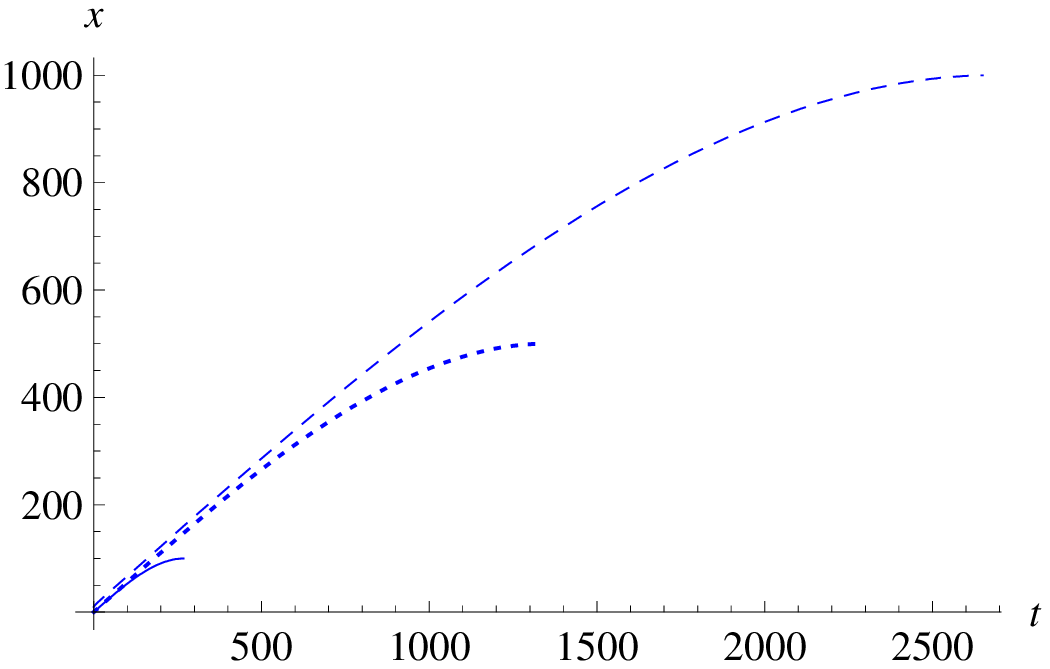}
\caption{Time evolution of the bubble in the outside coordinates
$x(t)$, and time evolution of the surface energy density
$\sigma(t)$. In these cases the static universe is dominated by dust
and the bubble wall contain dust. In all these graphics we have
considered dashed line for $R=1000$, dotted line for $R=500$ and
continuous line for $R=100$.\label{Soln1}}
\end{figure}


\begin{figure}
\centering
\includegraphics[width=7cm]{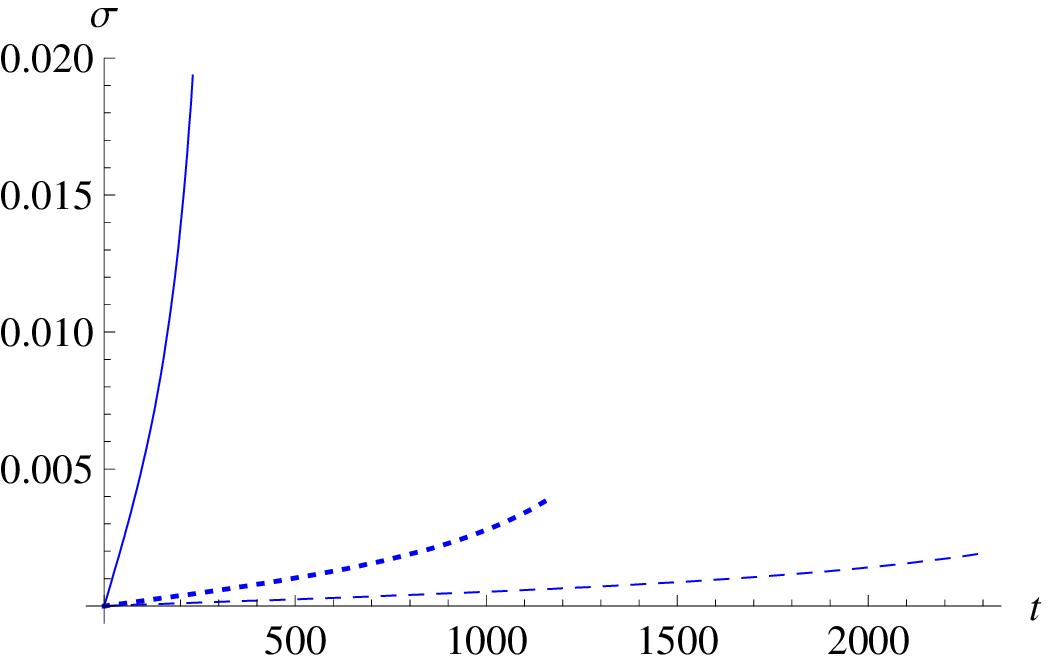}
\includegraphics[width=7cm]{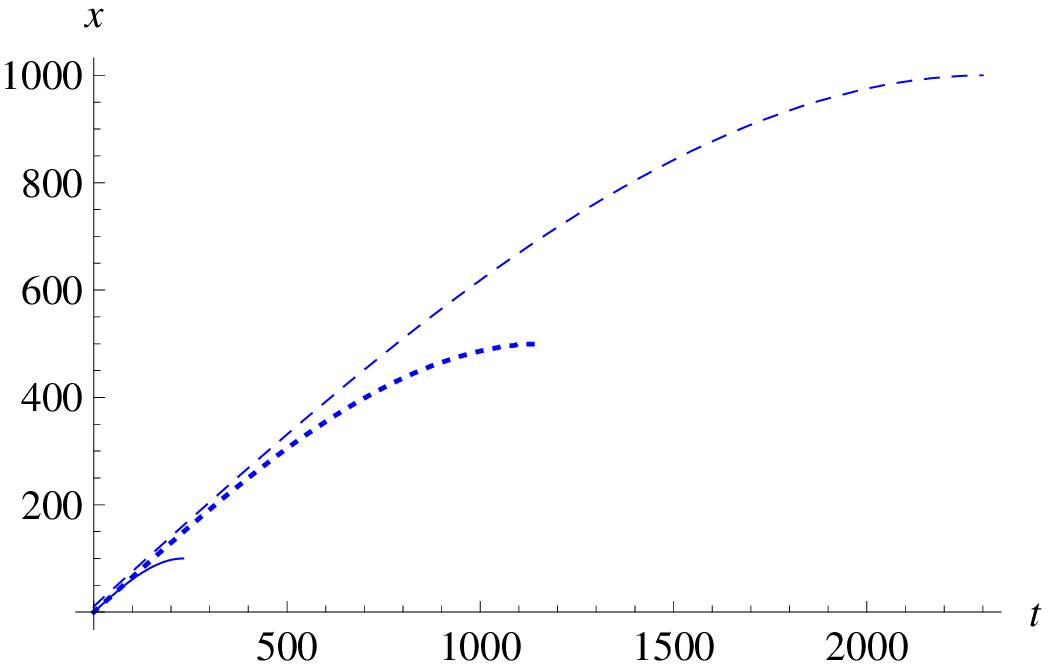}
\caption{Time evolution of the bubble in the outside coordinates
$x(t)$, and time evolution of the surface energy density
$\sigma(t)$. In these cases the static universe is dominated by
radiations and the bubble wall contain radiations.  In all these
graphics we have considered dashed line for $R=1000$, dotted line
for $R=500$ and continuous line for $R=100$.\label{Soln2}}
\end{figure}


\begin{figure}
\centering
\includegraphics[width=7cm]{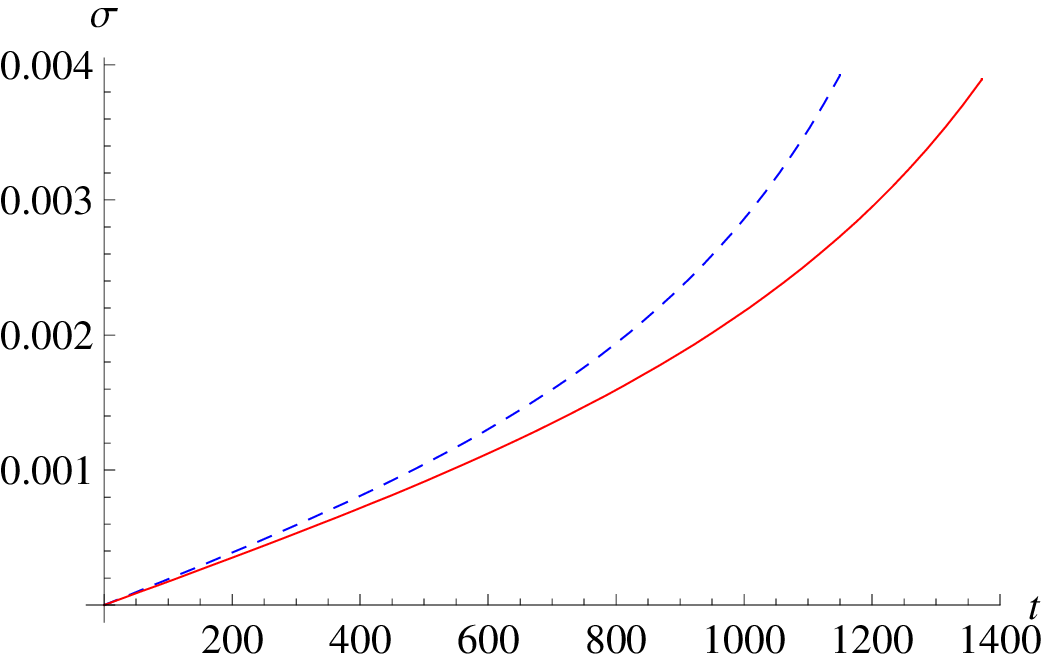}
\includegraphics[width=7cm]{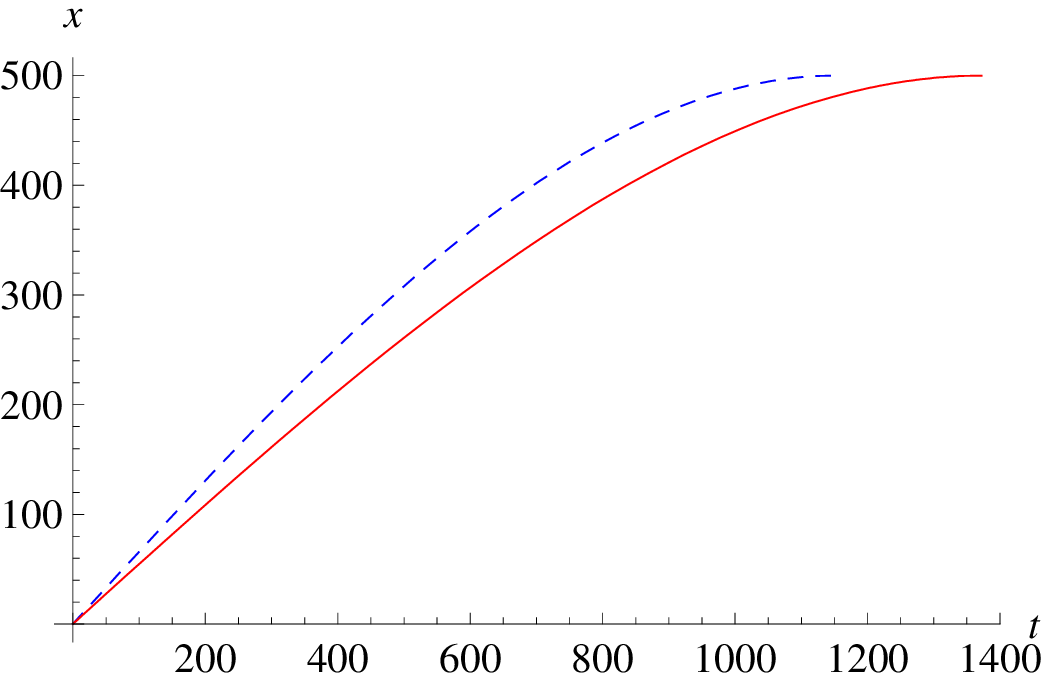}
\caption{Time evolution of the bubble in the outside coordinates
$x(t)$, and time evolution of the surface energy density
$\sigma(t)$, for a background with $R=500$. Dashed line corresponds
to a static universe dominated by dust and bubble wall containing
radiation. Continuous line corresponds to a static universe
dominated by radiation and a bubble wall containing
dust.\label{Soln3}}
\end{figure}

Some of the numerical  solutions are shown in
Figs.~(\ref{Soln1},\ref{Soln2}, \ref{Soln3}) where the evolution of
the bubble, as seen by the outside observer, is illustrated. In
these numerical solutions we have considered three different
curvature radius ($R=1000$, $R=500$, $R=100$) and various matter
contents combinations for the background and the bubble wall.
From these examples we can note that the bubble of the new face
grows to fill the background space, where the shell coordinate
asymptotically tends to the curvature radius $R$.

\section{Conclusions}\label{Sec-conclusions}

In this work we explore an alternative scheme for an Emergent
Universe scenario developed in \cite{Labrana:2011np}, where the
universe is initially  in a truly static state. This state is
supported by a scalar field which is located in a false vacuum. The
universe begins to evolve when, by quantum tunneling, the scalar
field decays into a state of true vacuum.

In particular, we study the process of tunneling of a scalar field
from the false vacuum to the true vacuum and the consequent creation
and evolution of a bubble of true vacuum in the background of
Einstein static universe.
The motivation in doing this is because we are interested in the
study of new ways of leaving the static period and begin the
inflationary regime in the context of Emergent Universe models.

In the first part, we study a Einstein static universe dominated by
two fluids, one is a standard perfect fluid and the other is a
scalar field located in a false vacuum. The requisites for obtain a
static universe under these conditions are discussed.
In the second part, we study the tunneling process of the scalar
field from the false vacuum to the true vacuum and the consequent
creation of a bubble of true vacuum in the background of Einstein
static universe. Following the formalism  presented in
\cite{KeskiVakkuri:1996gn} we found the semiclassical tunneling rate
for the nucleation  of the bubble in this curved space.
We conclude that the probability for the bubble nucleation is
enhanced by the effect of the curvature of the closed static
universe background.
In the third part of the paper, we study the evolution of the bubble
after its materialization. By following the formalism developed by
Israel \cite{Israel:1966rt} we found that once the bubble has
materialized in the background of an ES universe, it grows filling
completely the background space.
In particular, we use the approach of \cite{Fischler:2007sz} to find
the equations which govern the evolution of the bubble in the
background of the ES universe. These equations are solved
numerically, some of these solutions, concerning several type of
matter combinations for the background and the bubble wall, are
shown in Figs.~(\ref{Soln1}, \ref{Soln2}, \ref{Soln3}).

In resume we have found that this new mechanism for an Emergent
Universe is plausible and could be an interesting alternative  to
the realization of the Emergent Universe scenario.


We have postpone for future work the  study of this mechanism
applied to Emergent Universe based on alternative theories to
General Relativity, like Jordan-Brans-Dicke \cite{Jbd}, which
present stable past eternal static regime
\cite{delCampo:2007mp,delCampo:2009kp}.
It is interesting explore this possibility because Emergent Universe
models based on GR suffer from instabilities, associated with the
instability of the Einstein static universe \cite{Eddington}.
This instability is possible to cure by going away from GR, for
example, by consider a Jordan Brans Dicke theory, see
\cite{delCampo:2007mp, delCampo:2009kp}.
Another possibility is considering non-perturbative quantum
corrections of the Einstein field equations, either coming from a
semiclassical state in the framework of loop quantum gravity
\cite{Mulryne:2005ef,Nunes:2005ra} or braneworld cosmology with a
timelike extra dimension \cite{Lidsey:2006md, Banerjee:2007qi}. In
addition to this, consideration of the Starobinsky model, exotic
matter \cite{Mukherjee:2005zt, Mukherjee:2006ds} or the so-called
two measures field theories \cite{delCampo:2010kf,delCampo:2011mq,
Guendelman:2011fq,Guendelman:2011fr} also can provide a stable
initial state for the emergent universe scenario.

In the context of GR the instability of the ES could be overcome by
consider a static universe filled with a non-interacting mixture of
isotropic radiation and a ghost scalar field \cite{Barrow:2009sj} or
by consider a negative cosmological constant with a universe
dominated by a exotic fluid satisfies $P=(\gamma -1)\rho$ with $0 <
\gamma < 2/3$, see \cite{Graham:2011nb}.


\begin{theacknowledgments}

I would like to thank A. Andrianov, V. Andrianov, S. Afonin and all
the other colleagues at the University of Saint Petersburg for
organizing the II Russian-Spanish Congress where this work was
presented.
All figures in this review are taken from Phys.\ Rev.\ D {\bf 86},
083524 (2012). Copyright (2012) by The American Physical Society.
This work has been partially supported by FONDECYT grant N$^{0}$
11090410, Mecesup UBB0704 and Universidad del B\'{i}o-B\'{i}o
through grant DIUBB 121407 GI/VC.

\end{theacknowledgments}



\bibliographystyle{aipproc}   


\IfFileExists{\jobname.bbl}{}
 {\typeout{}
  \typeout{******************************************}
  \typeout{** Please run "bibtex \jobname" to optain}
  \typeout{** the bibliography and then re-run LaTeX}
  \typeout{** twice to fix the references!}
  \typeout{******************************************}
  \typeout{}
 }


\end{document}